\def\Msun{\hbox{$\rm\thinspace M_{\odot}$}}
\documentclass[useAMS,usenatbib]{mn2e}
\usepackage{graphicx}
\usepackage{txfonts}
\usepackage{natbib}

\title[]{X-ray reflection in a sample of X-ray bright Ultraluminous X-ray sources}

\author[M. D. Caballero-Garc\'{i}a \& A. C. Fabian]{M. D. Caballero-Garc\'{i}a$^{1}$\thanks{E-mail:
mcaballe@ast.cam.ac.uk} and A. C. Fabian$^{1}$ \\
$^{1}$Institute of Astronomy, Madingley Road, Cambridge, CB3 0HA}

\begin{document}

\date{}

\pagerange{\pageref{firstpage}--\pageref{lastpage}} \pubyear{2002}

\maketitle

\label{firstpage}

\begin{abstract}
  We apply a reflection-based model to the best available {\it
    XMM-Newton} spectra of X-ray bright UltraLuminous X-ray (ULX)
  sources (NGC~1313 X--1, NGC~1313 X--2, M~81 X--6, Holmberg~IX X--1,
  NGC~5408 X--1 and Holmberg~II X--1). A spectral drop is apparent in
  the data of all the sources at energies 6--7\,keV. The drop is
  interpreted here in terms of relativistically-blurred ionized
  reflection from the accretion disk.  A soft-excess is also detected
  from these sources (as usually found in the spectra of AGN), with
  emission from O K and Fe L, in the case of NGC~5408 X--1 and
  Holmberg~II X--1, which can be understood as features arising from
  reflection of the disk. Remarkably, ionized disk reflection and the
  associated powerlaw continuum provide a good description of the
  broad-band spectrum, including the soft-excess. There is no
  requirement for thermal emission from the inner disk in the
  description of the spectra. The black holes of these systems must
  then be highly spinning, with a spin close to the maximum rate of a
  maximal spinning black hole. The results require the action of
  strong light bending in these sources. We suggest that they could be
  strongly accreting black holes in which most of the
  energy is extracted from the flow magnetically and released above
  the disc thereby avoiding the conventional Eddington limit.

\end{abstract}


\section{Introduction}

Ultra-luminous X-ray (ULX) sources are point like, nonnuclear sources observed at other galaxies, with observed luminosities greater than
the Eddington luminosity for a $10\,{\rm M}_{\odot}$ stellar mass black hole (BH), with $L_{X}{\ge}10^{39}$\,${\rm erg\,s^{-1}}$ \citep{f1}. The true nature of these
objects is still open to debate \citep{mc1}. One fundamental issue is whether the emission is isotropic or beamed along our line-of-sight. A possible
scenario for geometrical beaming involves super-Eddington accretion during phases of thermal-timescale mass transfer \citep{k1}. Alternatively,
if the emission is isotropic and the Eddington limit is not violated, ULX must be fuelled by accretion onto Intermediate-Mass BH (IMBH), with masses
in the range 100-10\,000 ${\rm M}_{\odot}$ \citep{cm1}. Currently, there is no agreement regarding the nature of these sources.
It is possible that some ULX appear very luminous because of a combination of moderately high mass, mild beaming and
mild super-Eddington emission. It is also possible that ULX are an inhomogeneous population, comprised of both a subsample of IMBH and
moderately beamed stellar mass black holes \citep{f2,mc1}.

NGC~5408 X--1, HOLM~II X--1, M 81 X--9 (which is in the companion-dwarf galaxy Holmberg IX, hereafter called HOLM~IX X--1), the ULX in NGC~1313 and M 81~X--6 are ULX located in dwarf
(NGC~5408 X--1, HOLM~II X--1 and HOLM~IX X--1) and spiral (NGC~1313 X--1, NGC~1313 X--2 and M 81 X--6) galaxies, respectively. All these ULX peak in X-ray luminosity above $L_{X}=1{\times}10^{40}$\,${\rm erg\,s^{-1}}$,
thus being excellent targets for testing spectral models to the data. They are nearby and located at distances of $D=4.8,3.5,3.7,3.63,4.50$\,Mpc (\citealt{karachentsev02,stobbart06,liu08}), 
for NGC~5408 X--1, HOLM~IX X--1, NGC~1313 X--1, M~81 X--6 and HOLM~II X--1 respectively. NGC~5408 X--1 is among the few ULX for which (10-200\,mHz) Quasi-Periodic Oscillations
(QPO) have been found (\citealt{stroh1,dewangan06}). NGC~5408~X-1 exhibits X-ray timing and spectral properties analogous to those exhibited by Galactic stellar-mass black hole in the
{\it very high} or {\it steep power-law} state \citep{remi1}, but with the characteristic variability timescales (QPO and break frequencies) 
consistently scaled down \citep{stroh1}. For NGC~5408 X--1 the inferred characteristic size for the X-ray emitting region is ${\approx}$ 100 times larger than
the typical inner disk radii of Galactic black holes. The highest signal-to-noise spectra of ULX can often be fitted by composite models of a thermal disk together with a hard-powerlaw 
tail, with a low measured disk temperature of ${\approx}0.2$\,keV (\citealt{miller03,miller04}). If such factors are entirely due to a higher black hole mass, they imply an 
IMBH with $M{\approx}100-10\,000$\,${\rm M}_{\odot}$. Both the morphology and flux of the optical high-excitation nebulae detected around some ULX probably rule out strong beaming 
as the origin of the X-ray emission (\citealt{pakull03,kaaret04,soria1,abolmasov07}).
 
The detection for most ULX of spectral curvature, in the form of a deficit of photons at energies $E{\ge}2$\,keV (\citealt{roberts05,stobbart06,miyawaki09}) has led to the 
suggestion that most ULX have spectral properties that do not correspond to any of the accretion states known in Black Hole Binaries (BHB), making it unlikely that
ULX are powered by sub-Eddington flows onto an IMBH \citep{roberts07}. The application of Comptonization models to the data (\citealt{stobbart06,gladstone09}
and references therein) results in strikingly high and low values for the coronal opacity (${\tau}{\ge}5$) and the electron temperature ($kT_{e}=1-3$\,keV),
difficult to explain for the expected physical conditions in a corona surrounding the black hole.
This is very different to the typical values found for BHB during the {\it low/hard} state, with spectra dominated by Comptonization \footnote{It has to be noted 
that similar values, i.e. $kT_{e}=1-3$\,keV and ${\tau}{\ge}5$, have been found in the description of the spectra of BHB during the {\it steep powerlaw} state 
(e.g. GRO~J1655--40 and GRS~1915+105; \citealt{makishima00,kubota04,ueda09}).}, and appears irreconcilable with the IMBH model, which assumes that they operate as simple scaled-up BHB. 

Here, we present an alternative interpretation of the spectral shape, based on a physically-justified model commonly used on other accreting black holes.
The soft part of the spectrum ($E{\le}1$\,keV) -- the soft X-ray excess -- and the high-energy curvature are just aspects of a reflection spectrum expected from 
accretion (\citealt{guilbert88,george91}). A major component of reflection, the broad iron K line, has been found in many Seyfert galaxies (\citealt{tanaka95,nandra08}), 
accreting stellar-mass black holes (\citealt{miller07,reis09a}) and even accreting neutron stars (\citealt{cackett08,reis09b}). Both the soft excess and the relativistic
broad iron K line have recently been demonstrated to be part of the same physical process, i.e., the reaction of the disk to irradiation from a high-energy source, in the Seyfert-1
galaxy 1H 0707-495 \citep{fabian09}. In this paper we investigate whether reflection models can account for the spectra of these ULX and, if so, we put them in the context
of the accreting black holes known so far.

For this study, we choose the ULX with the best available data and with the longest exposure time observations (${\approx}100$\,ks of exposure time)
of the {\it XMM-Newton} satellite. By using only the highest quality data from the widest band pass, highest sensitivity instruments available we expect to make good statements 
on the accretion processes in these ULX. In Section \ref{observ} we describe the observations and data used, in Section \ref{spec_anal} we report on the results of the application
of fits with the reflection model to the spectra and in Section \ref{discuss} discuss the results obtained.

\section{Observations and data reduction} \label{observ}

In this work, we consider the time averaged {\it XMM-Newton} EPIC-pn spectra from the longest available observations of sources with a rate ${\ge}0.5$\,counts/s in the X-ray band. 
The datasets were obtained through the {\it XMM-Newton} public data archive. The EPIC-pn camera has a higher effective area than the EPIC-MOS cameras, and drives the results
of any joint spectral analysis. The reduction and analysis reported in this work used SAS version 8.0.0. We checked for pile-up in all 
the observations and found that this was not significant (i.e. less than 5 per cent for the high energy channels) for all the observations.

In Table \ref{log_obs} we present a log of the observations. We applied the standard time and flare filtering (rejecting high-background periods of rate ${\ge}0.4$\,counts/s, 
as recommended for the pn camera \footnote{Information provided at "node52.html" of the User Scientific Guide.}). We filtered the event files, 
selecting only the best-calibrated events (pattern${\le}4$ for the pn), and rejecting flagged events (flag$=0$).

For each exposure, we extracted the flux from a circular region on each source. The background was extracted from a 
circular region, not far from the sources. Both the coordinates of the centroids and the radius used for the regions were the same as those
employed by \citet{gladstone09}. We built response functions with the SAS tasks {\tt rmfgen} and {\tt arfgen}.
We fitted the background-subtracted spectra with standard models in XSPEC 12.5.0 \citep{a1}. All errors quoted in this work are $90\%$ confidence errors, obtained
by allowing all variable parameters to float during the error scan. Owing to the uncertainties in the EPIC calibration, we
used only the 0.3--10\,keV range. The resulting spectra were grouped with the FTOOL {\tt grppha} to bins with a minimum of 20 counts each.

\begin{table*}
 \centering
 \begin{minipage}{120mm}
  \caption{{\it XMM-Newton} observations log and effective exposure times of {\it XMM-Newton} instruments.}
  \label{log_obs}
  \begin{tabular}{@{}lcccc@{}}
  \hline
   Source          & Obs ID          & Date              & Exposure time (\,s)  &   Rate (counts\,${\rm s^{-1}}$) \\
 \hline
   NGC~1313 X--1      & 0405090101       &  2006-10-16     & 121\,212       &   0.73       \\
   NGC~1313 X--2      & 0405090101       &  2006-10-16     & 121\,212       &   0.67       \\
   HOLM~IX X--1       & 0200980101       &  2004-09-26     & 117\,023       &   1.54       \\
   NGC~5408 X--1      & 0302900101       &  2006-01-15     & 130\,549       &   0.96       \\
   M81~X--6           & 0111800101       &  2001-04-23     & 132\,663       &   0.48       \\
   HOLM~II X--1       & 0200470101       &  2004-04-15     & 104\,677       &   3.13       \\
\hline
\end{tabular}
\end{minipage}
\end{table*}

\section{Spectral Analysis and results} \label{spec_anal}

\begin{figure*}
 \includegraphics[bb=27 48 559 763,width=6cm,angle=270,clip]{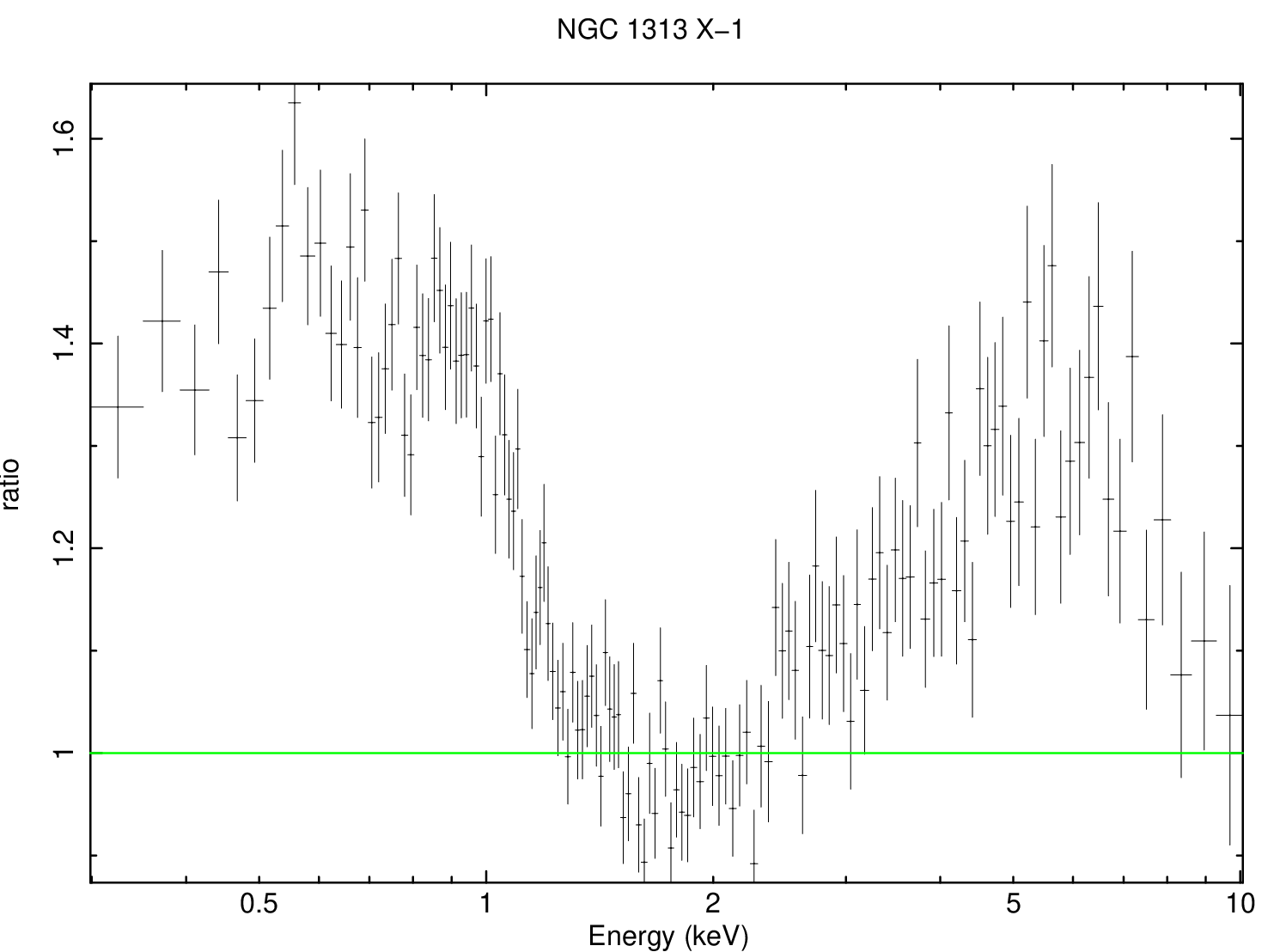}
 \includegraphics[bb=27 48 559 763,width=6cm,angle=270,clip]{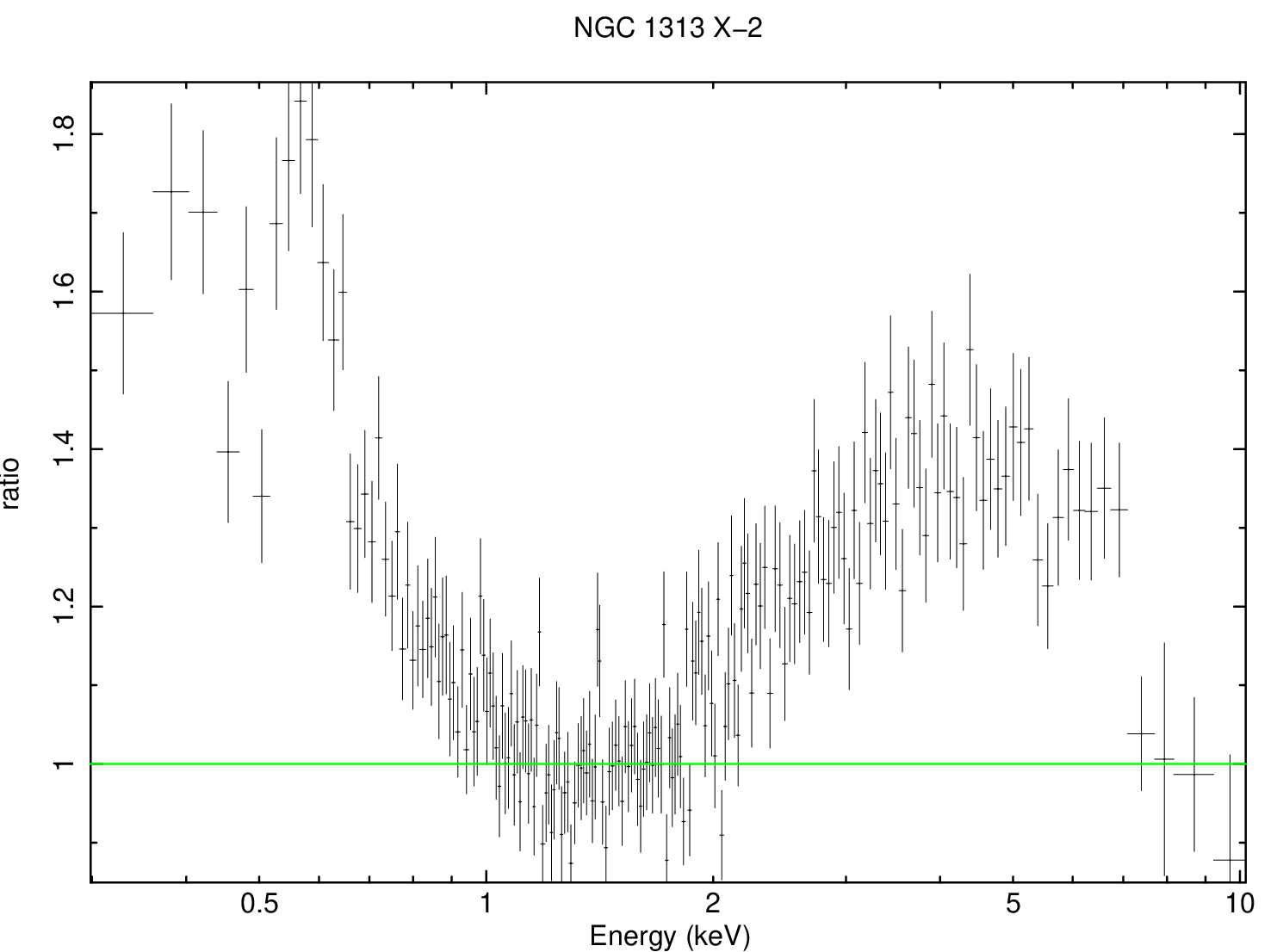}
 \includegraphics[bb=27 48 559 763,width=6cm,angle=270,clip]{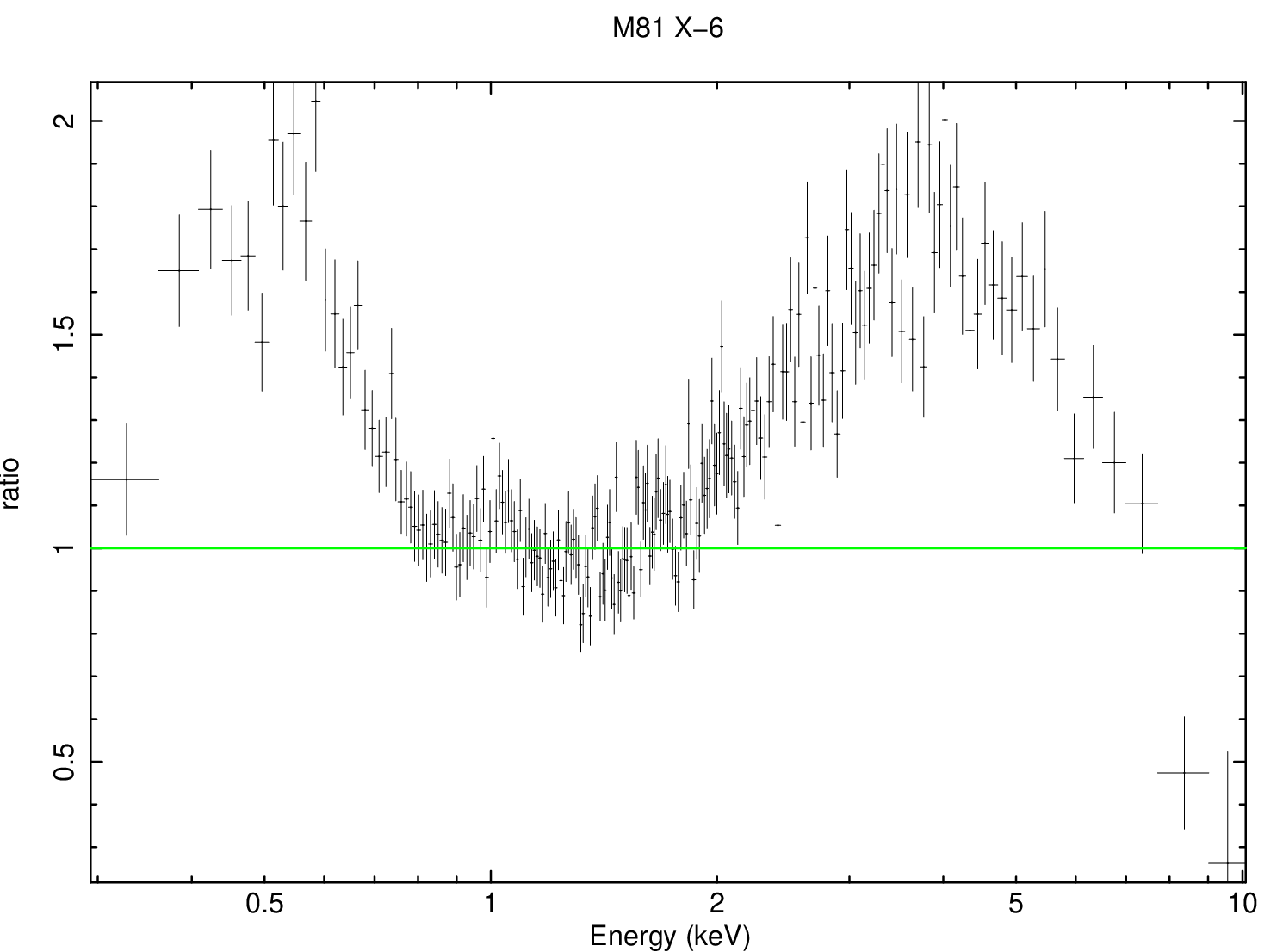}
 \includegraphics[bb=27 48 559 763,width=6cm,angle=270,clip]{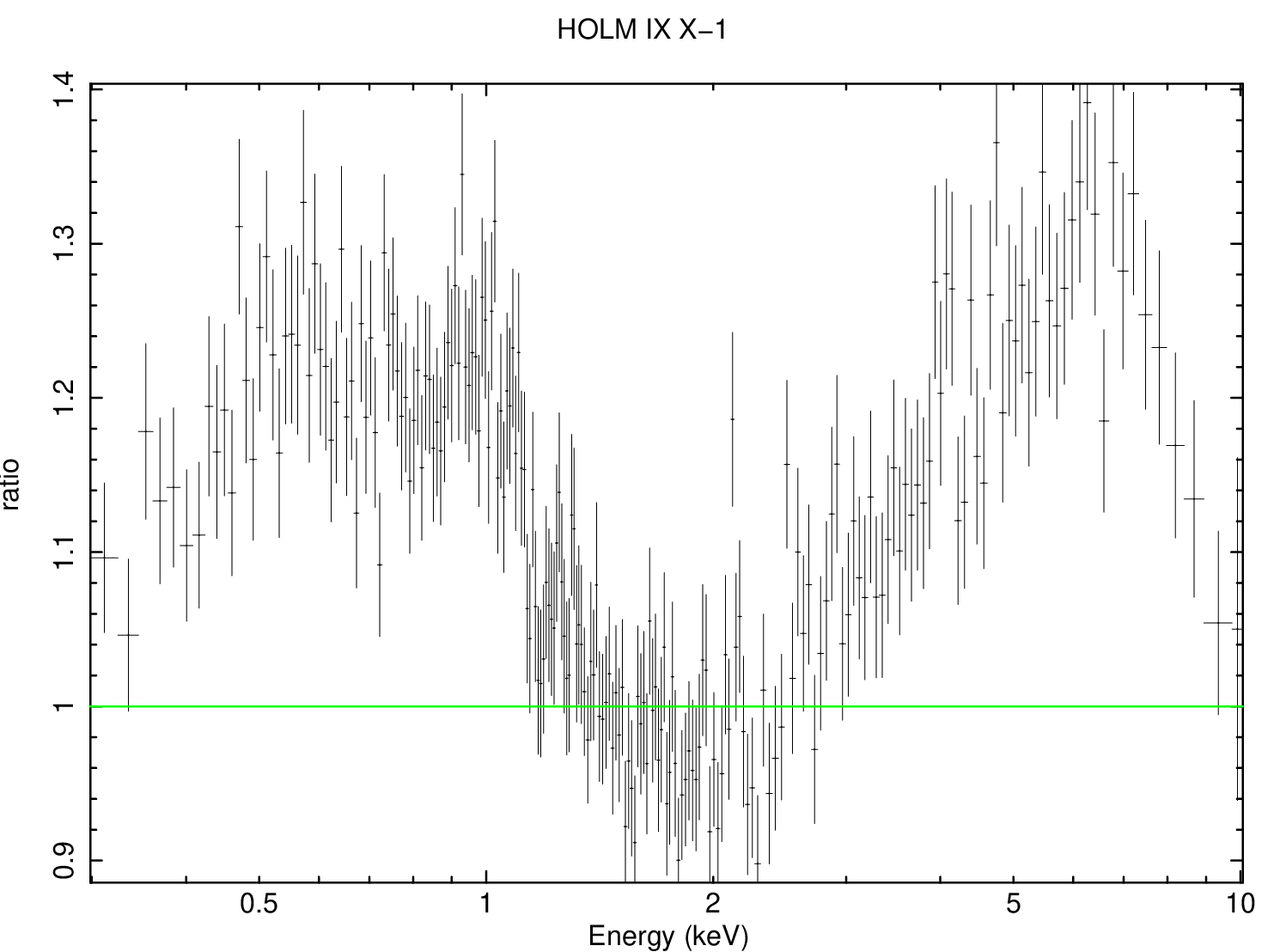}
 \includegraphics[bb=27 48 559 763,width=6cm,angle=270,clip]{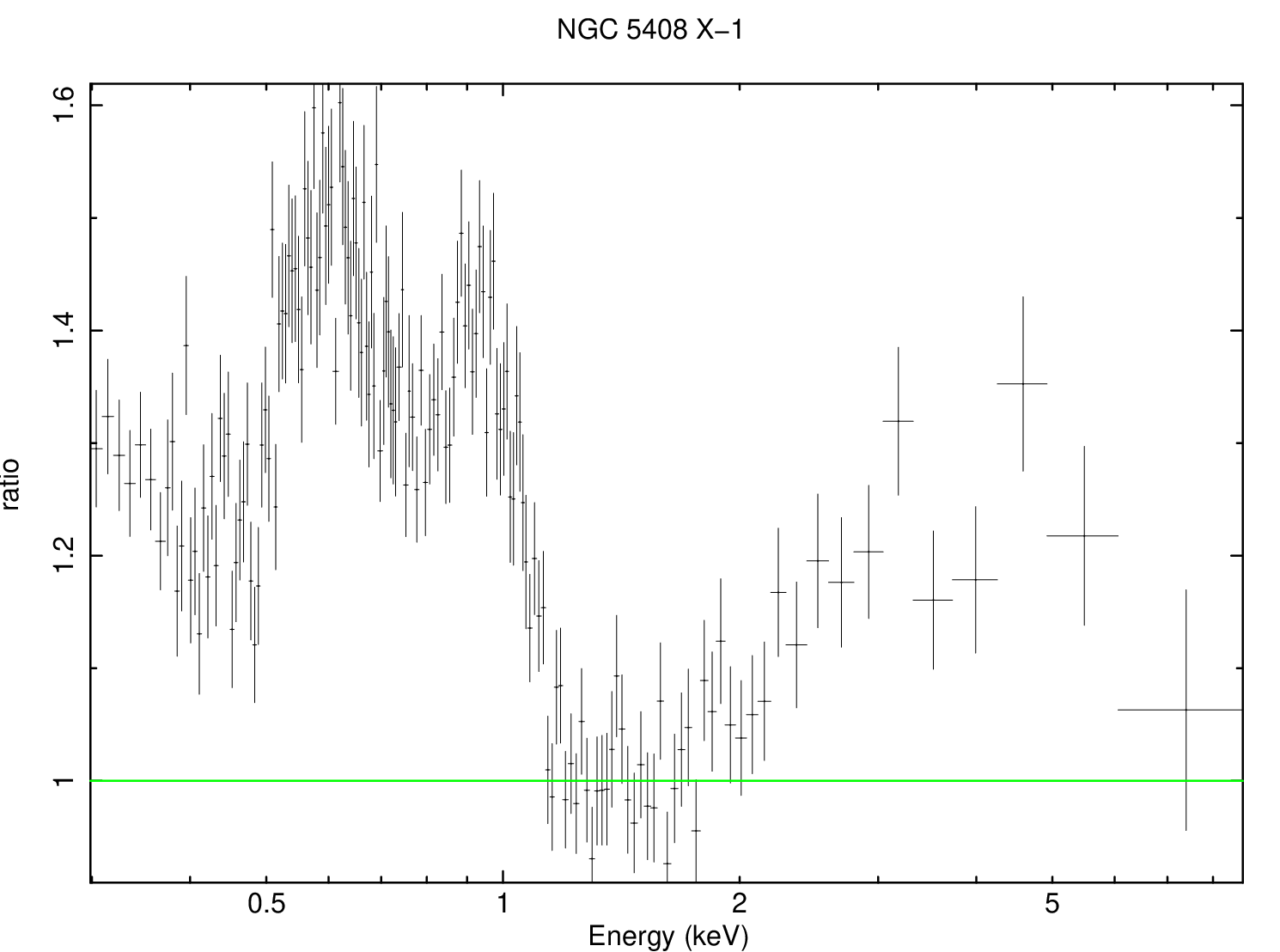}
 \includegraphics[bb=27 48 559 763,width=6cm,angle=270,clip]{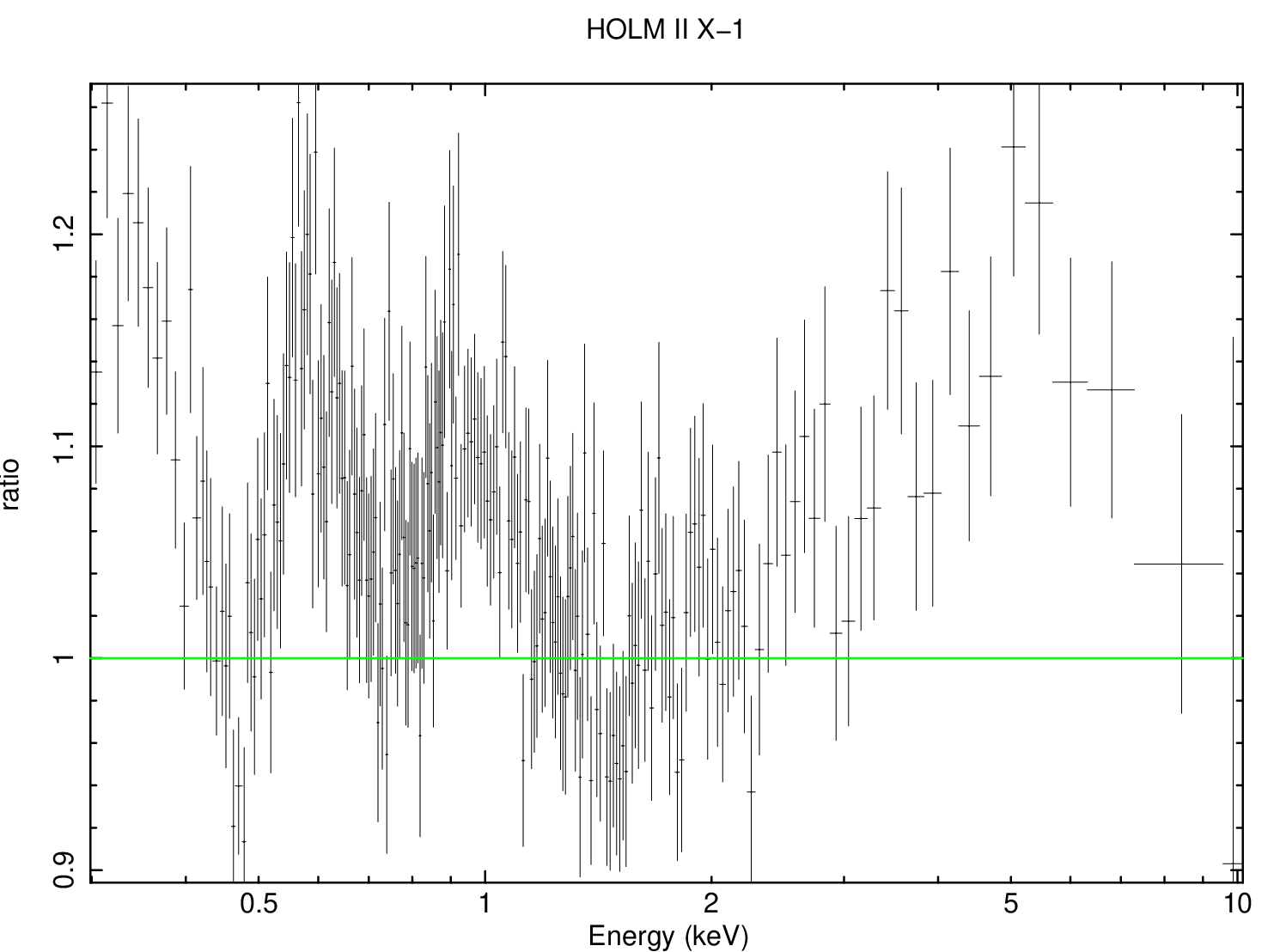}
 \caption{Ratio of the {\it XMM-Newton} pn spectra of NGC~1313 X--1, NGC~1313 X--2, M81~X--6, HOLM~IX X--1, NGC~5408 X--1 and HOLM~II X--1 (upper-left to lower-right) to the powerlaw which fits best between 1.2--3 and 8--10\,keV. 
A broad iron line and a soft excess are evident. Data have been 
rebinned for visual clarity.}
 \label{plots_ratio}
\end{figure*}

In contrast to previous work, where Comptonization models are used to fit the data, 
we looked at the observations of NGC~1313 X--1, NGC~1313 X--2, M81~X--6, HOLM~IX X--1, NGC~5408 X--1 and HOLM~II X--1 
from the point of view of a broad iron emission line, i.e. a reflection spectrum. An 
absorbed powerlaw was fitted to the data in the bands 1.2--3 and 8--10\,keV. In 
the case of NGC~1313 X--2 and M81~X--6 we used the energy range 1--2 and 8--10\,keV
instead. The ratio of the whole 0.3--10\,keV data set to the absorbed powerlaw is 
shown in Figure \ref{plots_ratio}. The column densities found are in the range of 
$N_{\rm H}=(0.1-0.5){\times}10^{22}$\,cm$^{-2}$,
thus larger than the corresponding galactic Hydrogen column densities \citep{dickey90}, 
indicating the presence of either local extra-absorption to the sources
or intrinsic to the host galaxies. A large skewed 
emission feature, similar to those found in BHB and AGN 
(\citealt{miller07,tanaka95,nandra08}) is evident about 3 and 8\,keV, and 
a soft excess (like those found in AGN) below 1\,keV. 

The data between 1.5 and 10\,keV were then modelled with a powerlaw
and a relativistic emission line \citep{laor91}. A good fit is
obtained with ${\Gamma}=1.5-3$ (see Table \ref{tparam}). 
The addition of this line represents an improvement in the statistics of 
${\Delta}{\chi}^{2}=60,115,367,103,65,10$ (for 5
extra d.o.f.) for NGC~1313 X--1, NGC~1313 X--2, M81~X--6, HOLM~IX
X--1, NGC~5408 X--1 and HOLM~II X--1, respectively. The line was thereafter
significant in all the sources with exception of HOLM~II X--1. HOLM~II X--1 has
the lowest {\tt laor} line detection, due to both the short exposure
time and steepness of the spectrum (we will refer to this point
later). Further longer exposure observations of HOLM~II X--1 are needed to fully address
this point. Notice that no extreme inclination is required to reproduce
the spectral drop (${\approx}40-60^{\circ}$). The strength of the
line (equivalent width), the rest frame energy and the emissivity
index of the line are in the range of $EW=0.15-1.27$\,keV,
$E=6.4-6.97$\,keV and $q=6-10$ (see Table \ref{tparam}).  For the case
of NGC~1313 X--1, NGC~1313 X--2, M81 X--6 and HOLM~IX X--1 the
rest-frame energies imply that we are dealing with ionized reflectors
(the steepness of the powerlaw for the case of NGC~5408 X--1 and
HOLM~II X--1 did not allow us to obtain reliable values for the line
parameters). There are remarkable emission line spectral
features around $E{\approx}0.6$ and $1$\,keV in the spectra of
NGC~5408 X--1 and HOLM~II X--1 (see Figure \ref{plots_ratio}), which
seem likely to come from O VIII K and Fe XVII and XVIII L.  These
lines have been usually understood as coming from the hot diffuse gas
around these sources and/or emission from a photoionized stellar wind,
similar to what is seen in some high-mass X-ray binaries in our own
Galaxy (\citealt{feng05,stobbart06}). \citet{goad06} argued that these
features commonly seen in the spectra from ULX correspond to poorly
modeled spectra of ULX considering absorbers with solar metal
abundances (i.e. they instead found that the RGS spectrum of HOLM~II
X--1 is consistent with an absorber of 0.6 times solar
abundance). Nevertheless, this value does not mean anything about the
local environment of the source, since this is the metal abundance
found in the total galactic column density in the direction to the
source \footnote{In this paper we apply a model describing the X-ray
  emission local to the source which incorporates the metal abundances
  as a parameter, thus potentially allow us to unveil the
  metallicities local to these sources.}. As pointed by
\citet{ross05}, these lines can plausibly come from the disk of the
sources. They originate through X-ray ionized reflection, which
occurs when a surface is irradiated with X-rays. The reflection
continuum is then complex and needs to be modelled properly.

\subsection{Reflection spectra}

We have therefore fitted \citet{ross05} ionized reflection models to
the whole 0.3--10\,keV observable band. Very good agreement is
obtained with a model which is relativistically blurred, has
supersolar abundances between 4 and 10 and photon index of
${\Gamma}{\approx}1.5-1.6$ for NGC~1313 X--1, NGC~1313 X--2 and
HOLM~IX X--1, ${\Gamma}{\approx}2$ for M81~X--6 and steeper
(${\Gamma}{\ge}2.5$) for NGC~5408 X--1 and HOLM~II X--1 (see Table
\ref{tparam}). In the case of NGC~5408 X--1 and HOLM~II X--1, this
model has worse statistics (${\chi}^{2}/{\nu}=736/577,945/779$),
presumably due to the steepness of the spectrum, which makes the
statistics be dominated by the softest part of the spectrum
($E{\le}1.5$\,keV). This would make easy the determination of the O
abundance (because the lowest part of the spectrum is very well
sampled) at the expense of having Fe badly constrained (being
responsible for the emission at the highest part of the
spectrum). Nevertheless, a study with variable abundances is beyond
the scope of this paper.  This model provided a good description of
the data (see Table \ref{tparam}). The ionization parameter of the
reflection model is in the range of ${\xi}=29-2\,500$. The
high values found for the ionization parameter for NGC~1313 X--1,
NGC~1313 X--2, M~81 X--6 and HOLM~IX X--1 are very likely related to
the featureless low energy ($E{\le}1$\,keV) spectra for these sources.

The relativistic blurring is accomplished using a Kerr kernel
\citep{laor91} with a radial emissivity index in the range of $q=4-9$
(where the emissivity has the radial dependence $r^{-q}$) from $R_{\rm
  in}=1.2-1.4$\,$R_{g}$ for NGC~1313 X--1, NGC~1313 X--2 and M81~X--6
and $R_{\rm in}=2.0-2.5$\,$R_{g}$ for HOLM~IX X--1, NGC~5408 X--1 and
HOLM~II X--1.  Here, 1$R_{g}=GM/c^{2}$. The outermost radius is fixed
to $400$\,$R_{g}$, a canonical value inferred from fits to spectra of
AGNs.  Since there is no previous measurement for the inner disk
inclination so far, it was constrained in the fits to be within the
range of $i=40-60^{\circ}$, i.e. an intermediate value. The inner disk
radii found for the former ones indicate that the matter reaches the
last stable orbit for a very rapidly spinning black hole, as inferred
from the highly skewed profiles for the broad Fe K line for these
sources (see Figure \ref{plots_ratio}). They show very strong
reflection components as well ($F_{\rm refl}/F_{\rm total}=0.7-1$),
i.e.  the spectrum is dominated by reflection, thus being {\it
  reflection-dominated} sources. Indeed, NGC~1313 X--2 and M81~X--6
are the most extreme cases, with no signs of emission from a powerlaw
high-energy source. In contrast, for the cases of HOLM~IX X--1,
NGC~5408 X--1 and HOLM~II X--1, the spectra are powerlaw dominated
({\it powerlaw-dominated} sources). Curiously, the reflection
dominated sources are the ones which show the smallest inner disk
radii, perhaps indicating that light bending is an important effect
for them.  It is worth noting that for the most {\it
  reflection-dominated} sources, i.e. NGC~1313 X--2 and M81~X--6, the
residuals at the broad Fe complex look peaky and that this is a
natural consequence of the fact that we are modelling with a uniform
disk, which could differ in properties to its outer region \footnote{The 
emissivity parameter is expected to vary between the
  inner and the outer boundaries of the disk, giving very smooth (high
  $q$) and peaky (low $q$) components, respectively. In this paper we
  just model the smooth component, expected to come from the 
  inner region of the disk.}. In contrast, for the powerlaw dominated
sources, the inner disk radii are larger. In the case of HOLM~IX X--1,
the larger radius ($R_{\rm in}=2.0$\,$R_{g}$) is accompanied by the
flattest powerlaw in the overall sample (${\Gamma}=1.504{\pm}0.004$),
and this could imply that reflection occurs farther from the black
hole with respect to the reflection dominated ones. In the case of
NGC~5408 X--1 and HOLM~II X--1, with steeper powerlaw indices
(${\Gamma}{\ge}2.5$) and showing worse fit statistics, due to the
lower signal-to-noise of the high-energy channels, no definite
statements can be made.  In none of the spectra does the addition of a
disk component ({\tt diskbb} in XSPEC) improve the
statistics. Therefore, we conclude that we cannot detect thermal
emission from the disk for these sources (we shall return to this
point in Section \ref{discuss}).

The fits with both the {\tt laor} and the reflection models above
indicate high values of the emissivity parameter ($q=4-9$), indicating
that relativistic blurring is an important effect and the emission is
concentrated near the innermost radii. The respective inner disk radii
obtained are low ($R_{in}=1.25-2.5\,R_{g}$), indicating X-ray emission
from regions very close to the black hole, where strong light bending
dominates. These effects mean that high values of the spins of the
black holes are expected. To study this possibility further, we
substituted the Kerr kernel ({\tt kdblur}), which assumes maximal
spin, by the relativistic kernel of ({\tt kerrconv};
\citealt{brenneman06}), which assumes variable spin parameter
($a=cJ/GM^{2}$; where $J$ is the angular momentum of a black hole of
mass $M$). We assumed the inner disk radii to be equal to the marginal
stability radius (i.e. the closest inner disk radius to the black
hole) and constrained the disk inclination to an intermediate value
($i=40-60^{\circ}$). Thanks to the flat spectra of NGC~1313 X--1,
NGC~1313 X--2, M~81 X--6 and HOLM~IX X--1, which lead to a strong signal at
high energies, we could constrain well the values of the emissivity
parameter of the disk and the spin of the black hole. The spins for
the black holes of these sources are very high, and close to the value
of a maximal spinning black-hole ($a=0.998$), with M~81 X--6 hosting
the most rapidly spinning black hole of the sample
($a=0.993-0.998$). In the case of HOLM~II X--1, the lowest value found
($a=0.47{\pm}0.12$) might be a consequence of the steep powerlaw and
short exposure, which lead to little high-energy signal (NGC~5408 X--1
is similar). Future work is needed in order to fully address this
point. As in the application of the Kerr kernel, the values obtained
for the ionization parameters are higher for NGC~1313 X--1, NGC~1313
X--2, M~81 X--6, HOLM~IX X--1 than in NGC~5408 X--1, HOLM~II X--1,
which correspond to featureless low-energy spectra for the former ones
due to the current exposure times used. The opposite occurs for
NGC~5408 X--1 and HOLM~II X--1, for which we could constrain very well
the ionization parameter (i.e. the disk density) to
${\xi}=49.4{\pm}0.8,29{\pm}6$, due to the good signal of the spectrum
at low energies ($E{\le}1.5$\,keV). This value is low for both sources
and indicates a high density for the reflecting medium (i.e. the
disk).

For all the sources, we find that iron is overabundant compared to the
Sun by $4-9$ times. Usually, ULX have been associated with metal poor
environments, because stars in these sites lose less mass during their
evolution and can reach very high masses and collapse to high-mass
black holes ($M{\le}80$\,${\rm M}_{\odot}$) at the end of their lives
\citep{fryer01}. Furthermore, there is observational evidence of real
low-metal abundances in the galaxies which host these ULX. Metal
abundance measurements from the H\,II regions in NGC~1313 and HOLM~II
X--1 have values $Z{\approx}0.008,0.1$ (\citealt{hadfield07,pakull02})
and \citet{mapelli09} have found similar values for other ULX host
galaxies. These facts agree with the known mass-metallicity relation,
in which the smallest galaxies are the poorest in metals.
\citet{winter07} found metal overabundance for a sample of ULX, with
mean metallicity close to the solar value \footnote{The sources with
  the longest exposure times in their observations, i.e.  HOLM~II X--1
  and HOLM~IX X--1 clearly are outliers of this relation.}.
With exception of HOLM~II X--1 and HOLM~IX X--1, where the metal 
abundance they obtain is compatible with ours, we can explain the discrepancy
in the level of metals by means of the different model used for the
soft excess (thermal emission from an accretion disk in their
case). Nevertheless, they found, like in our study, that the galaxies
do not obey the mass-metallicity relation in X-rays. This does not
mean a conflict with optical measurements, since we are measuring
metallicities local to the sources in X-rays.  One possible scenario
to explain such high values for the metallicities observed local to
the sources, without any apparent chemical metal enrichment in their
surrounding environments, is that the metals were blown out by the
type II SNe event and then recaptured (similar to the fallback disk
model of \citealt{li03}). We shall return to this point later in
Section \ref{discuss}.

\begin{figure*}
   \centering
 \includegraphics[bb=25 49 539 748,angle=270,width=8cm,clip]{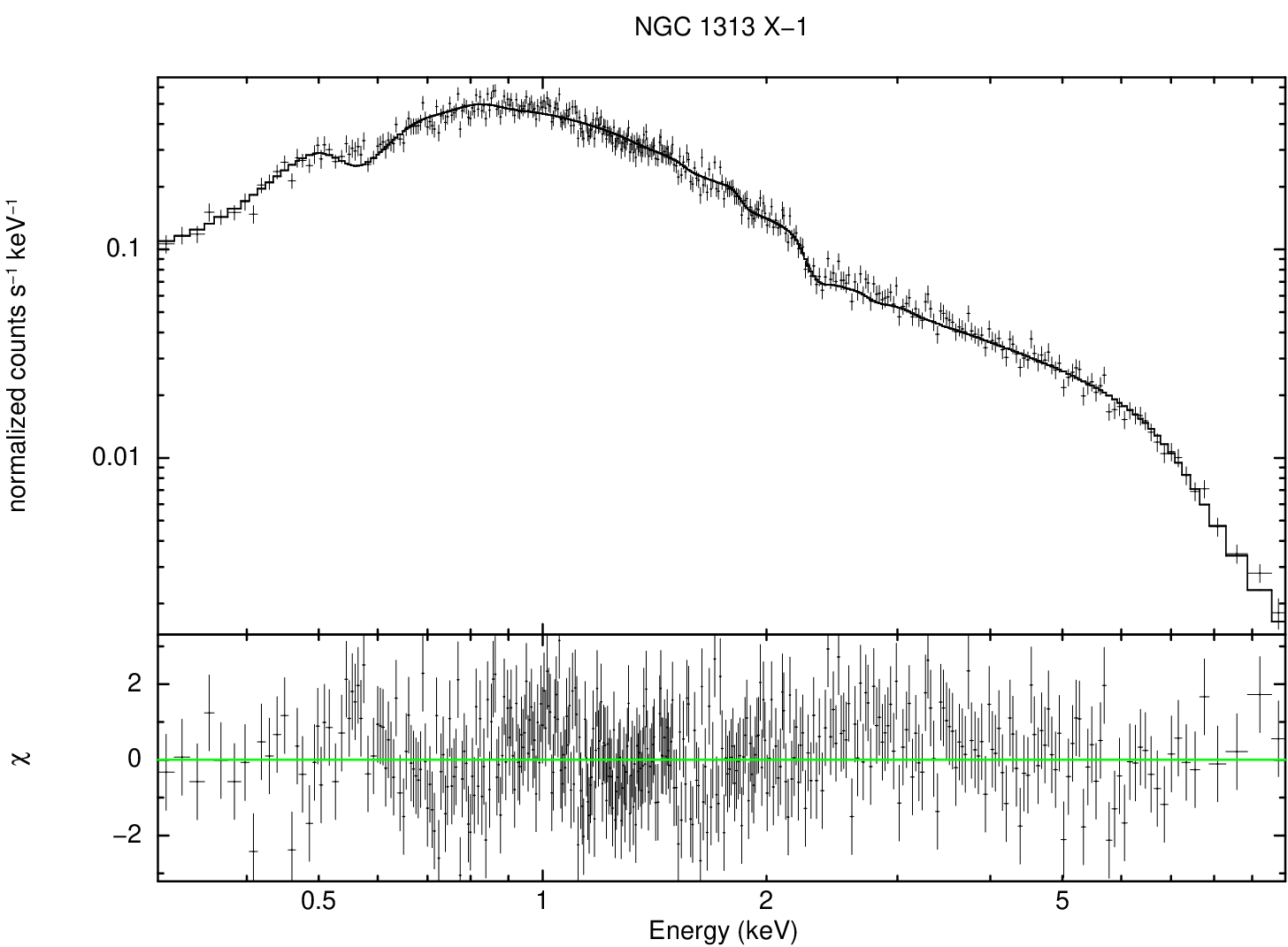}
 \includegraphics[bb=25 49 539 748,angle=270,width=8cm,clip]{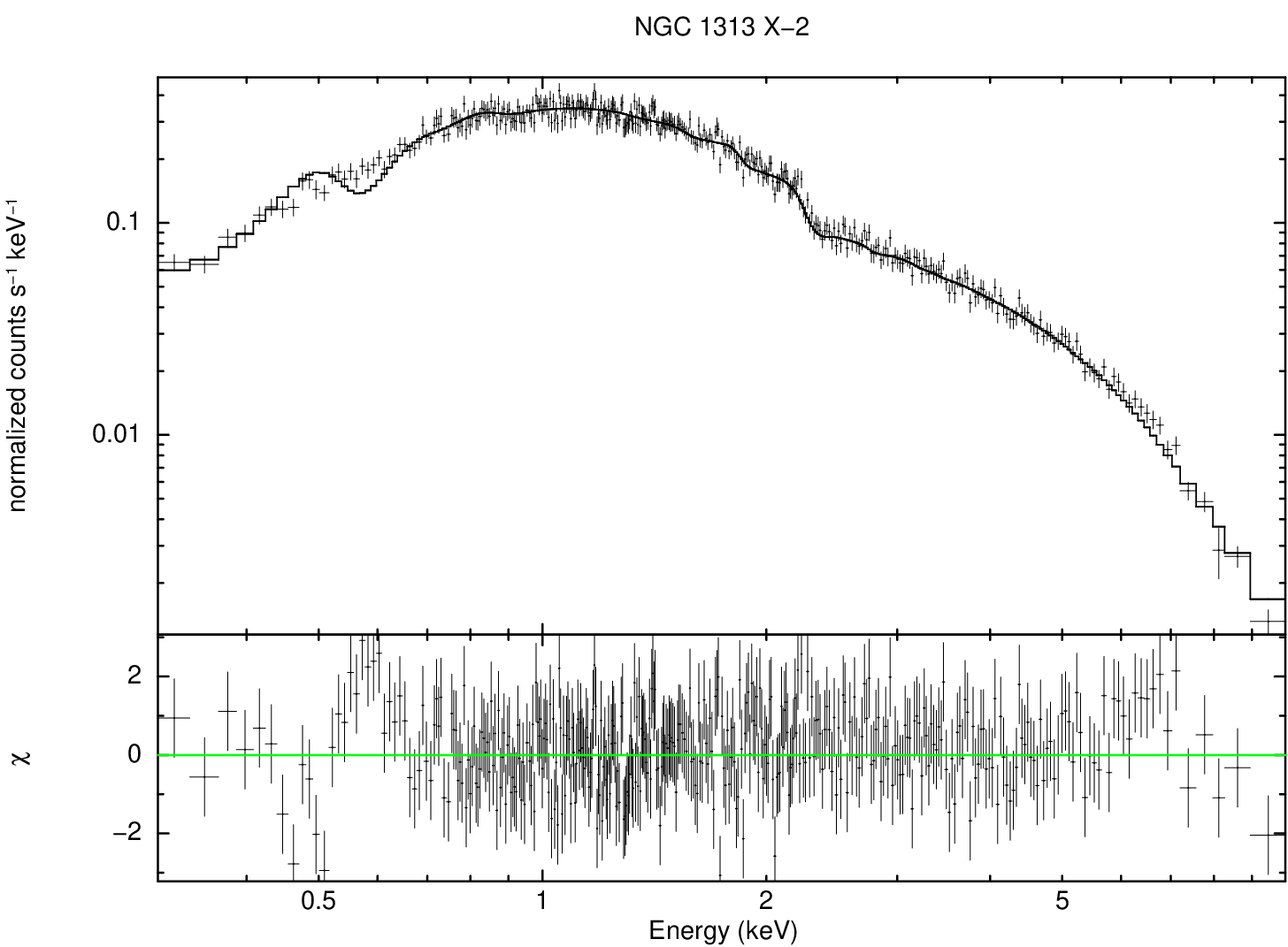}
 \includegraphics[bb=25 49 539 748,angle=270,width=8cm,clip]{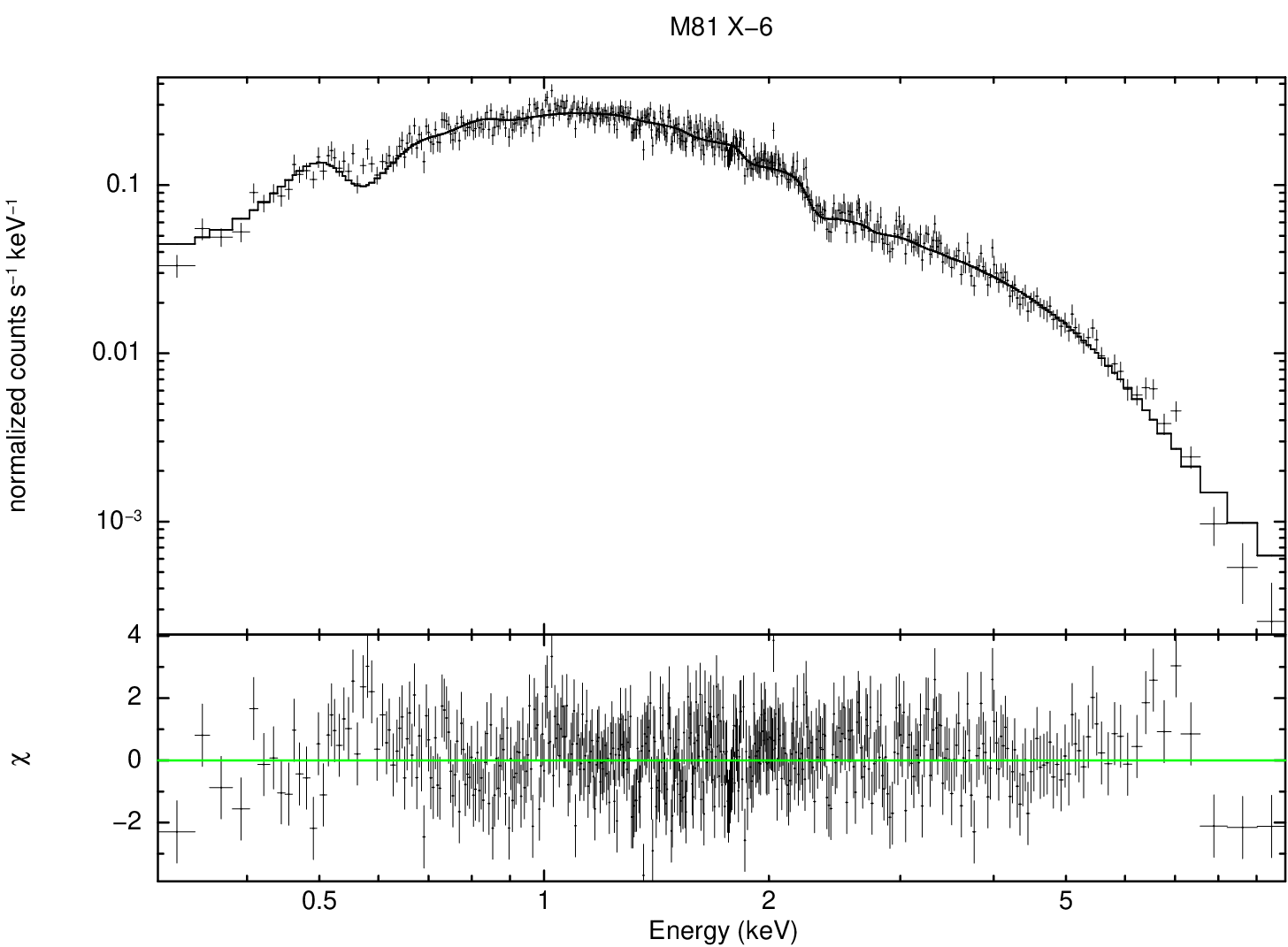}
 \includegraphics[bb=25 49 539 748,angle=270,width=8cm,clip]{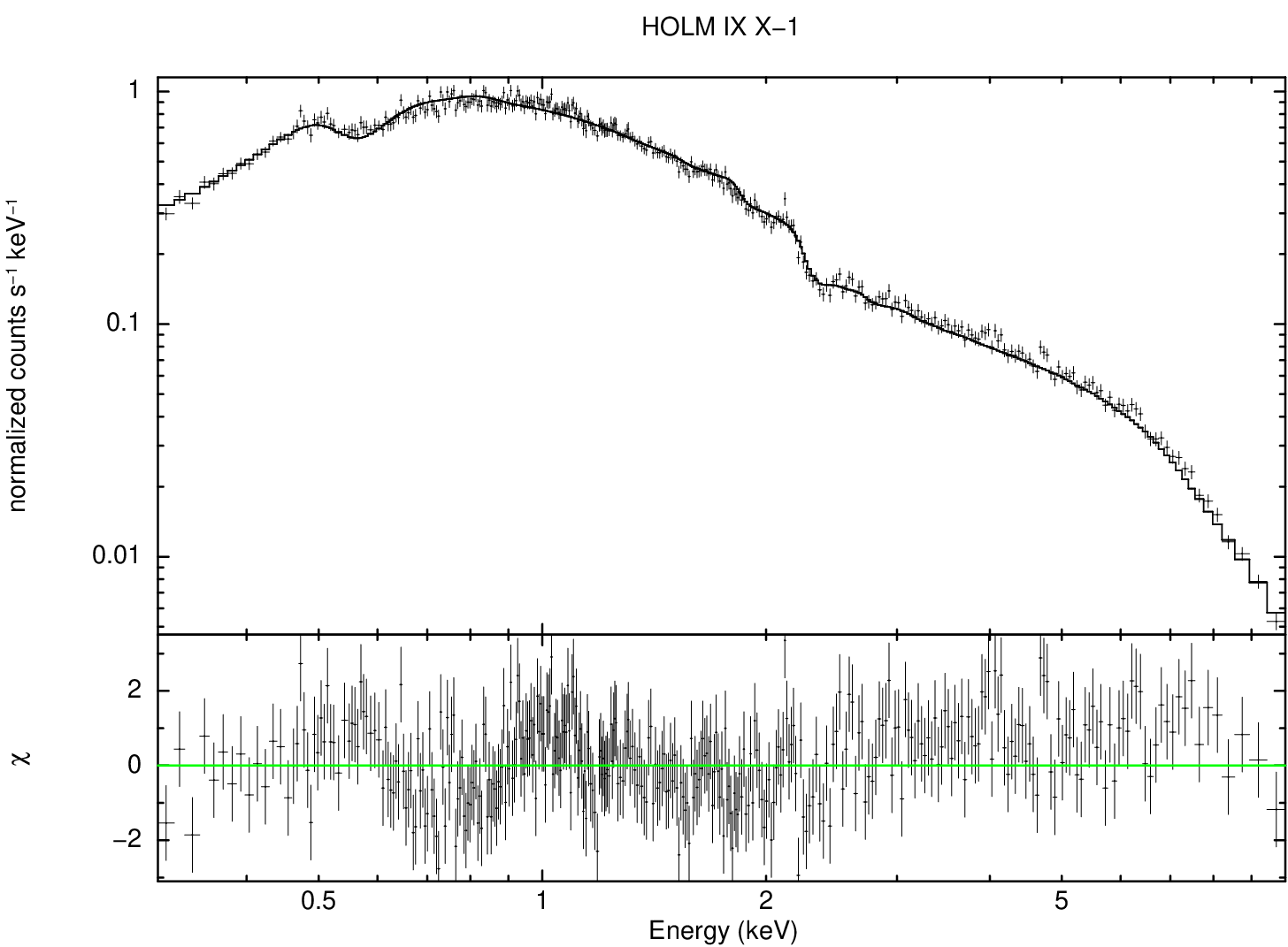}
 \includegraphics[bb=25 49 539 748,angle=270,width=8cm,clip]{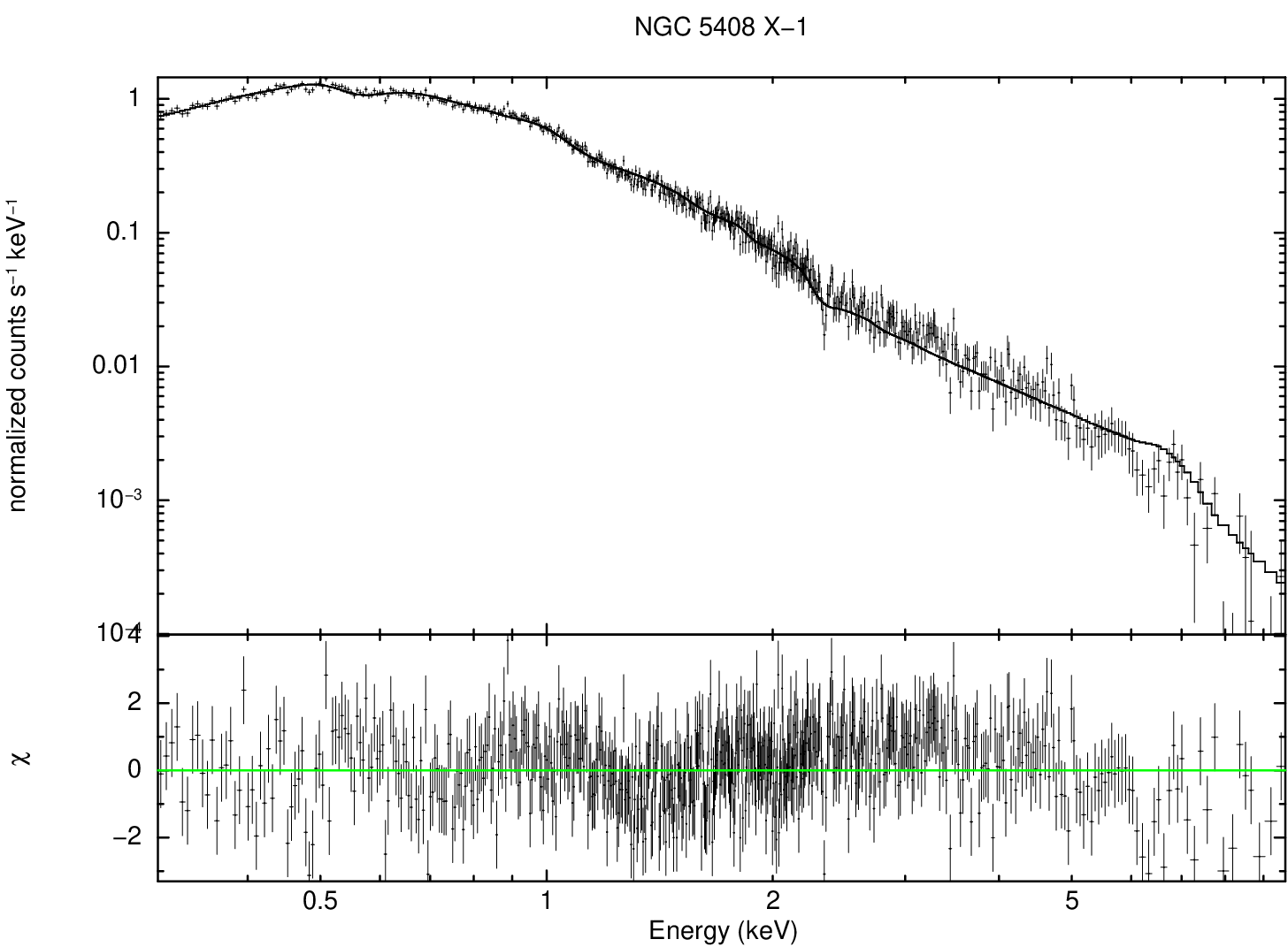}
 \includegraphics[bb=25 49 539 748,angle=270,width=8cm,clip]{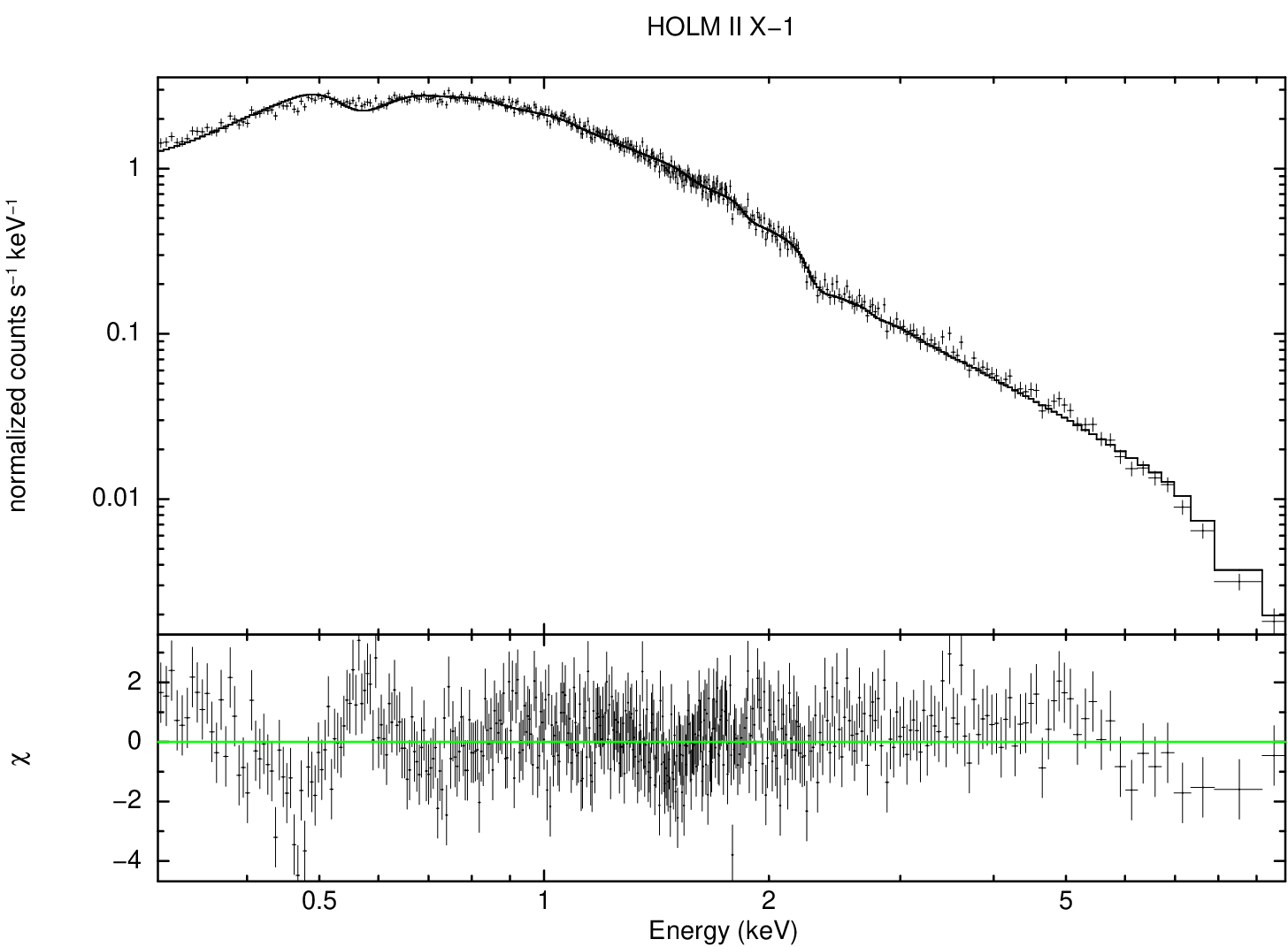}
 \caption{The best-fitting (blurred) ionized reflection plus power-law model plotted on the spectra of NGC~1313 X--1, NGC~1313 X--2, M~81 X--6, HOLM~IX X--1, NGC~5408 X--1 and HOLM~II X--1 (upper-left to lower-right). Simple photoelectric absorption has also been applied. The model fits the data very well over the entire observed energy band. Data have been rebinned for visual clarity.}
 \label{plots_reflion_data}
\end{figure*}

\begin{figure*}
   \centering
 \includegraphics[bb=28 44 559 769,angle=270,width=8cm,clip]{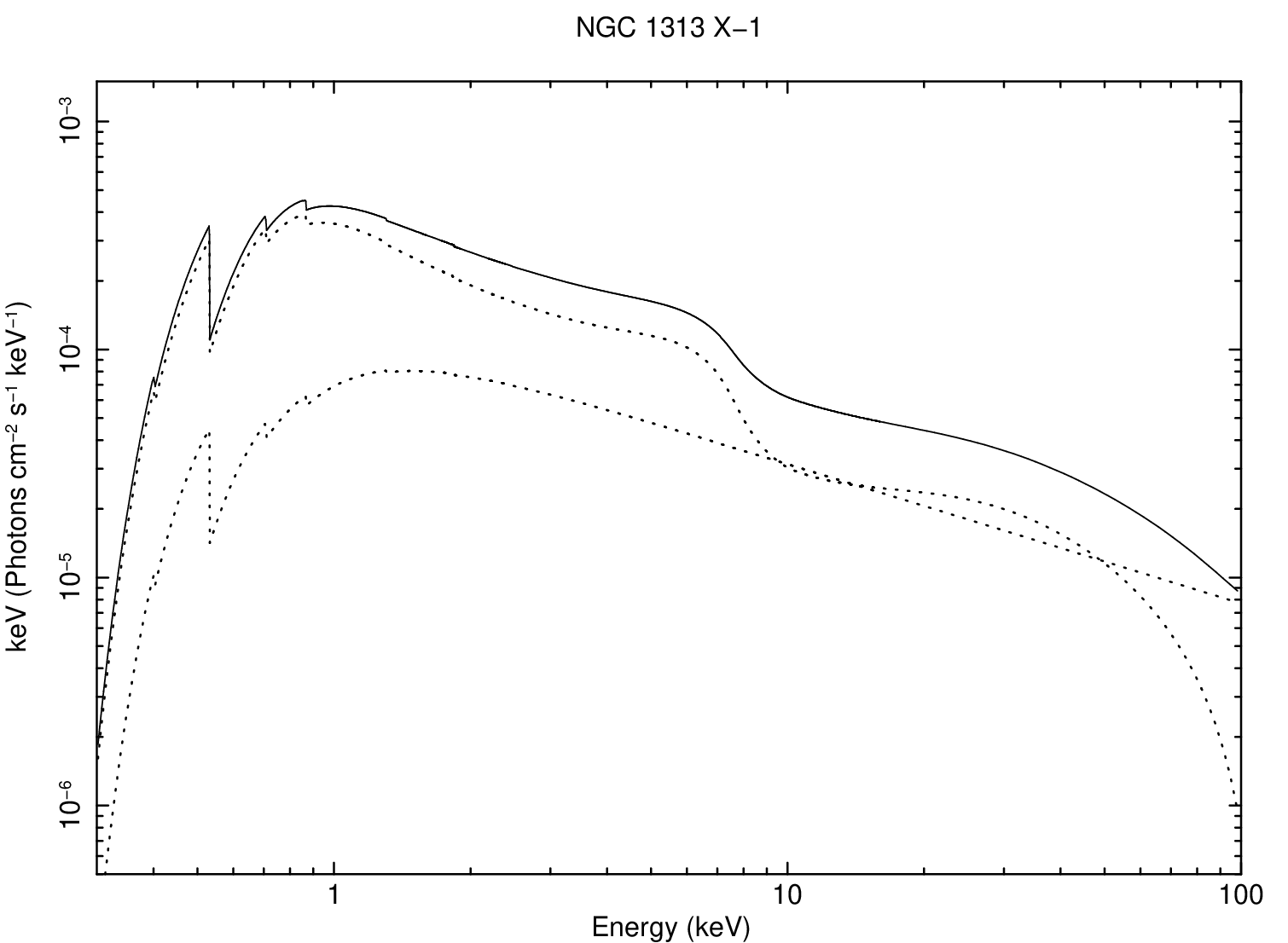}
 \includegraphics[bb=28 44 559 769,angle=270,width=8cm,clip]{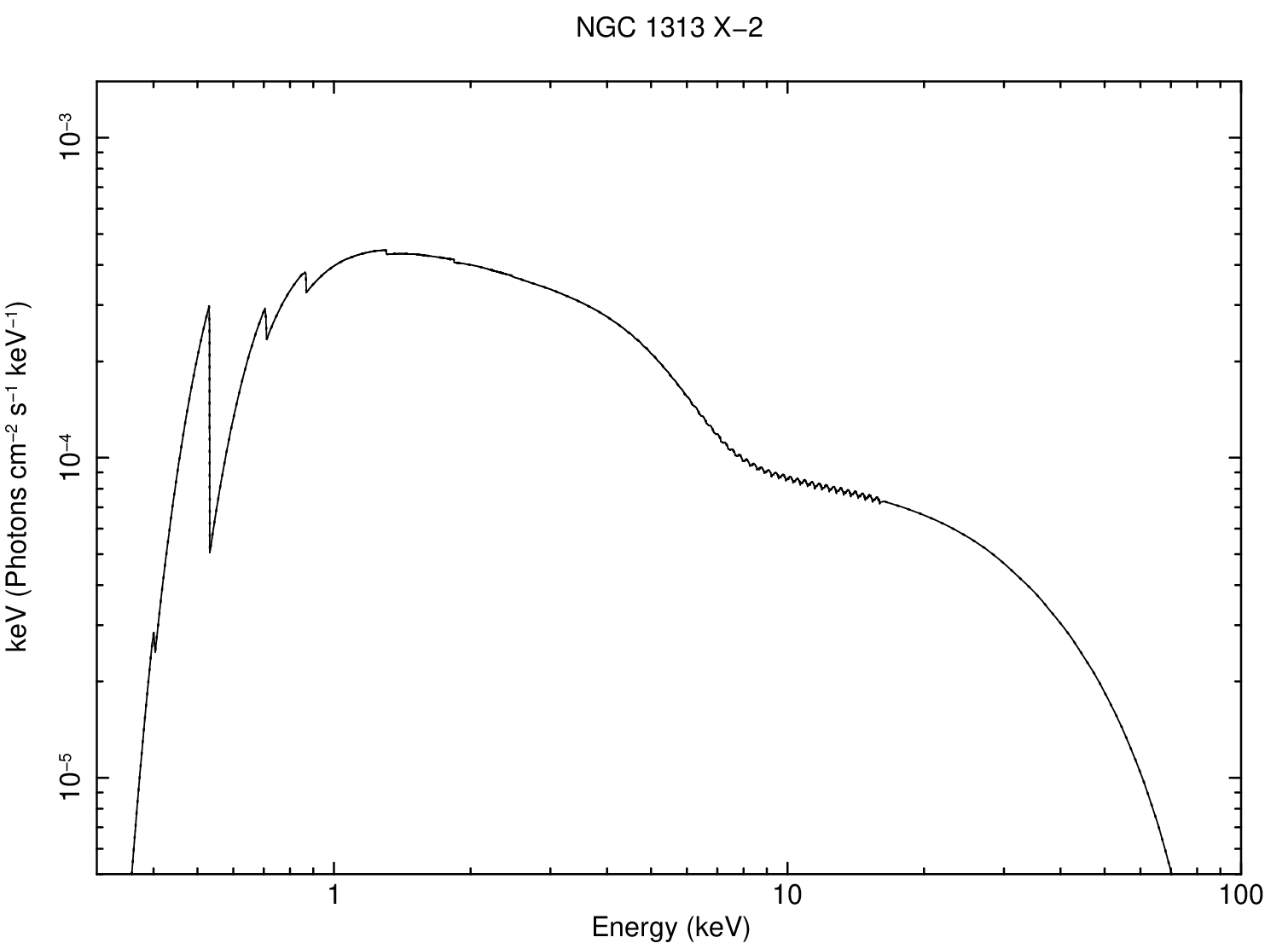}
 \includegraphics[bb=28 44 559 769,angle=270,width=8cm,clip]{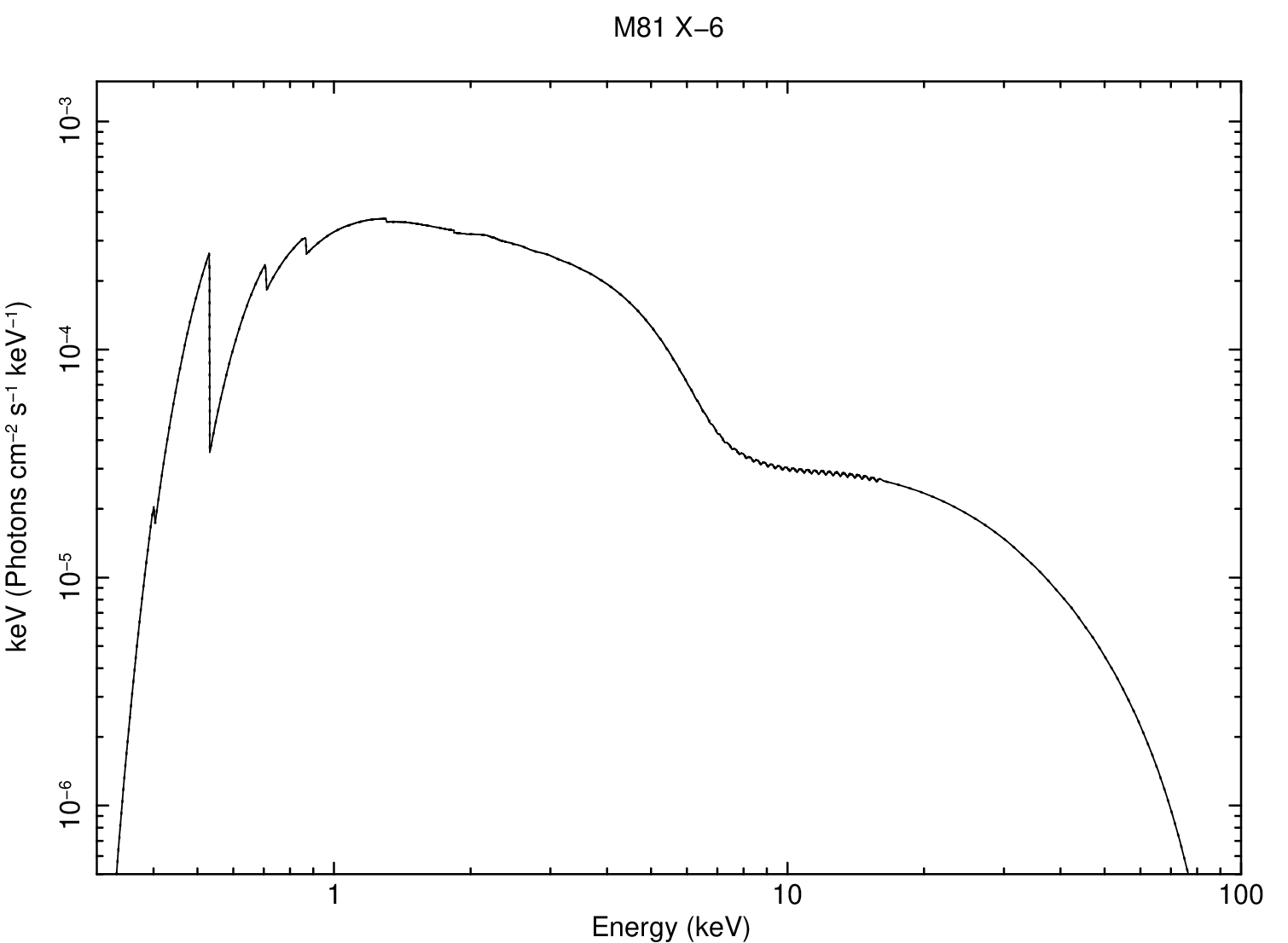}
 \includegraphics[bb=28 44 559 769,angle=270,width=8cm,clip]{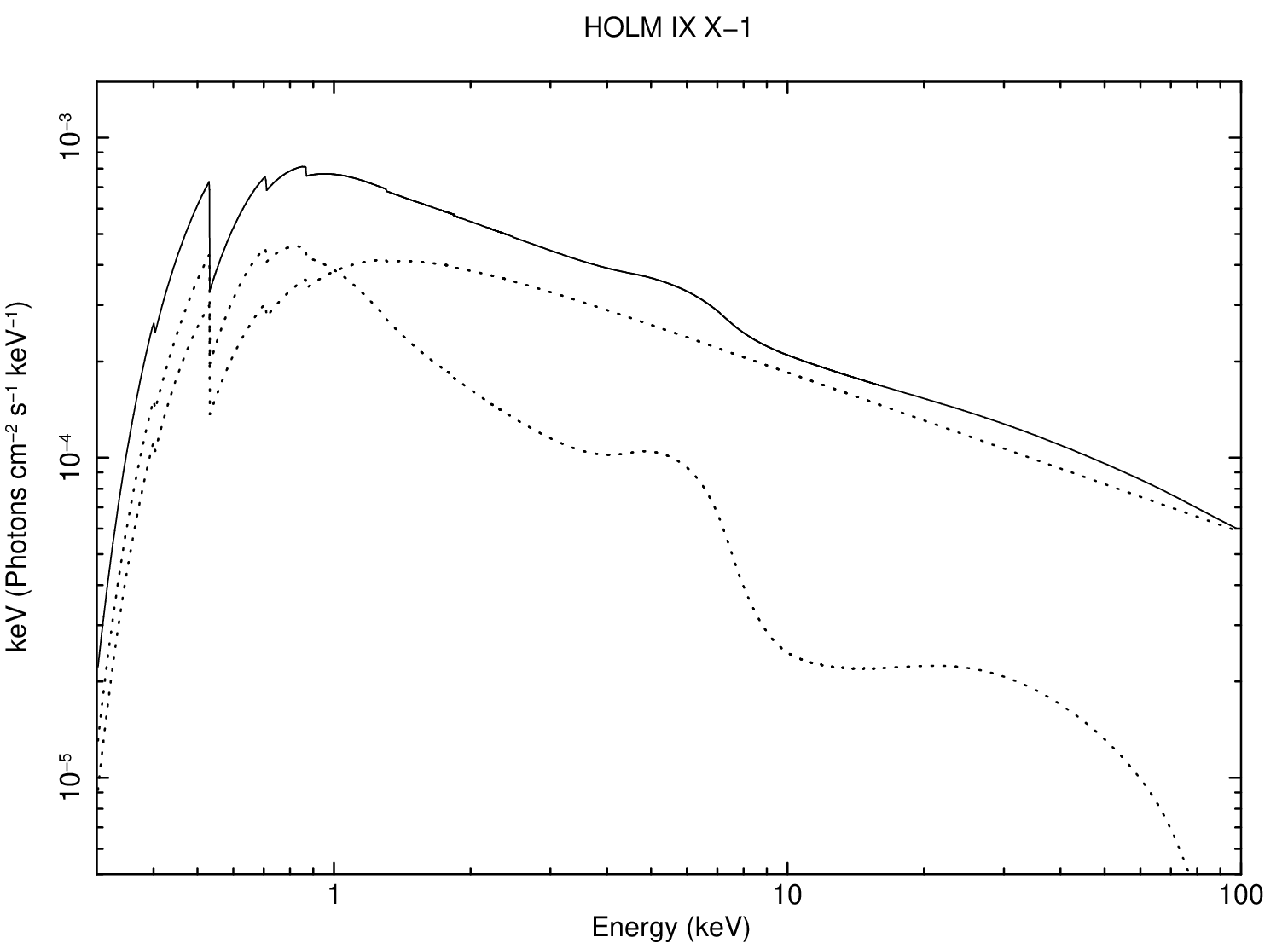}
 \includegraphics[bb=28 44 559 769,angle=270,width=8cm,clip]{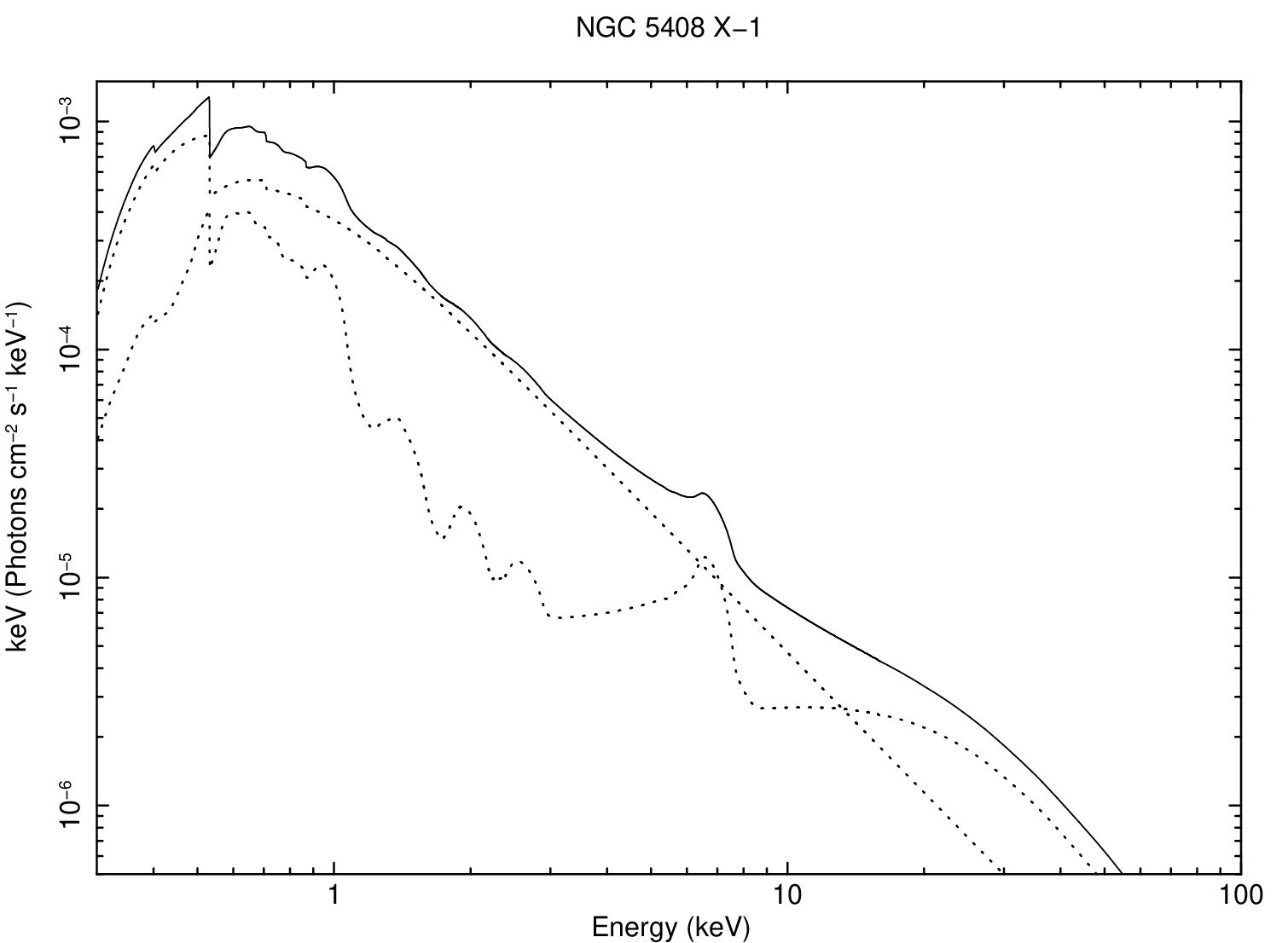}
 \includegraphics[bb=28 44 559 769,angle=270,width=8cm,clip]{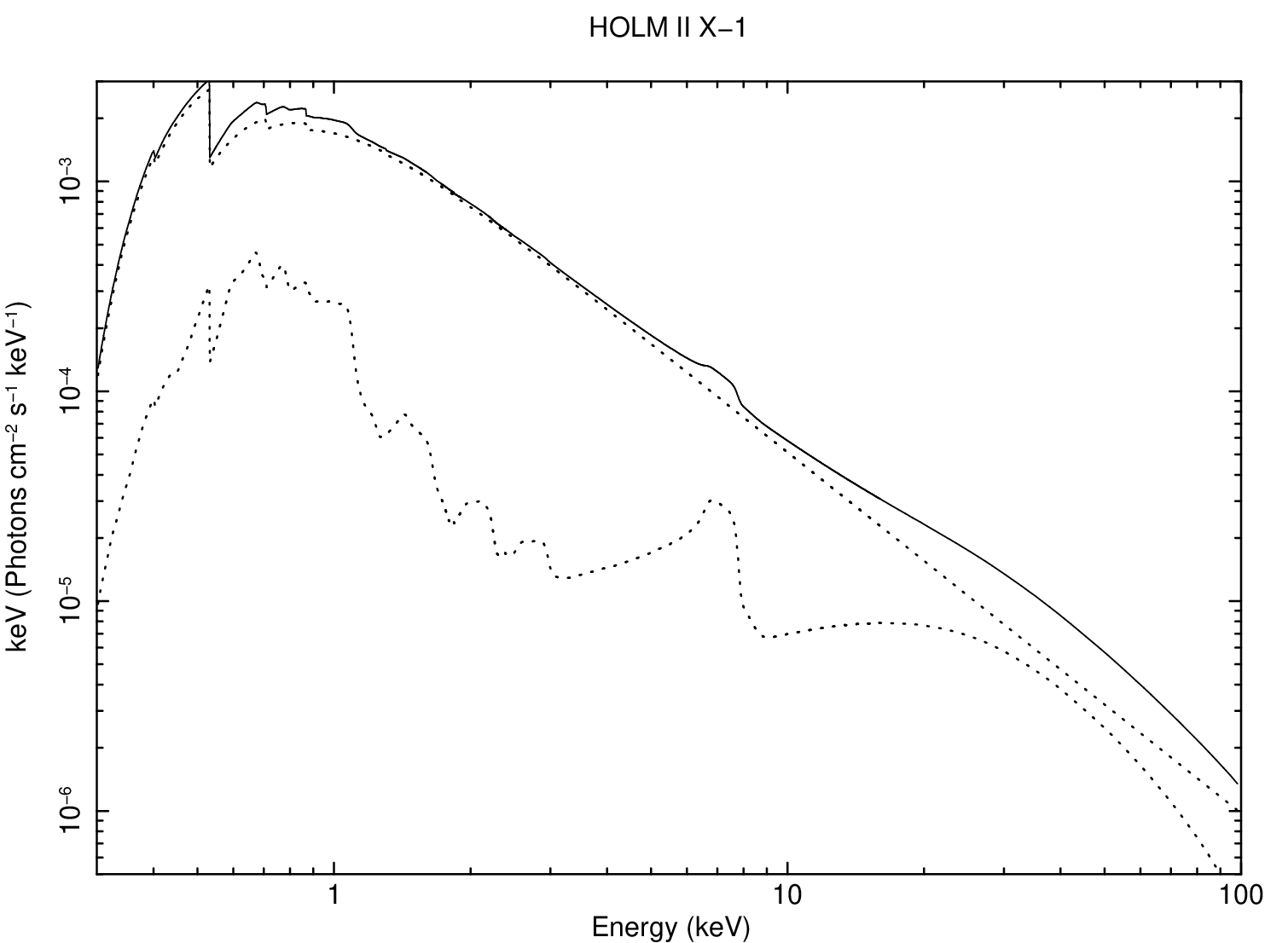}
 \caption{The (blurred) ionized reflection plus power-law model for NGC~1313 X--1, NGC~1313 X--2, M~81 X--6, HOLM~IX X--1, NGC~5408 X--1 and HOLM~II X--1 (upper-left to lower-right) used in the fits shown in Figure \ref{plots_reflion_data}. }
 \label{plots_reflion_model}
\end{figure*}

\begin{table*}
 \centering
 \begin{minipage}{166mm}
  \caption{Spectral parameters from fits with the {\tt laor} and the reflection model.}
  \label{tparam}
  \begin{tabular}{@{}lcccccc@{}}
  \hline
   Parameter                              &   NGC~1313 X--1                 &    NGC~1313 X--2     &    M~81 X--6         &  HOLM~IX X--1      &  NGC~5408 X--1    & HOLM~II X--1                            \\
 \hline
                      &                                 &   Laor Model                                     &                                     \\
\hline
 $E$\,(keV)                               &    $6.79{\pm}0.18$              &   $6.87{\pm}0.10$    &   $6.85{\pm}0.13$    &  $6.91{\pm}0.06$     &    $6.48{\pm}0.08$        &   $6.4{\pm}0.7$                          \\
 ${\rm EW}$\,(keV)                        &    $0.36$                       &   $0.45$             &   $1.27$             &  $0.70$              &    $0.43$                 &   $0.15$                                   \\
 $q$                                      &    $7.1{\pm}0.6$                &   $9.1{\pm}0.5$      &   $9.95{\pm}0.05$    &  $6.8{\pm}0.5$       &    $9.77{\pm}0.23$        &   $8.05{\pm}1.9$                          \\
 ${\Gamma}$                               &    $1.77{\pm}0.03$              &   $2.02{\pm}0.05$    &   $2.56{\pm}0.02$    &  $1.57{\pm}0.02$     &    $2.94{\pm}0.09$        &   $2.61{\pm}0.03$                          \\
 ${\chi}^{2}/{\nu}$                       &    $645/680$                    &   $709/711$          &   $646/599$          &  $1\,000/1\,036$     &    $380/335$              &   $578/539$                              \\
\hline
                      &                                 &   Blurred Reflection Model    &                                     \\
\hline
 $N_{H}$\,$(10^{22}$\,${\rm cm^{-2}})$    &    $0.241{\pm}0.002$            &   $0.374{\pm}0.003$  &   $0.422{\pm}0.004$  &  $0.164{\pm}0.002$   &    $0.129{\pm}0.001$    &   $0.179{\pm}0.003$                       \\
 ${\Gamma}$                               &    $1.627{\pm}0.013$            &   $1.504{\pm}0.005$  &   $2.043{\pm}0.012$  &  $1.504{\pm}0.004$   &    $3.04{\pm}0.01$      &   $2.72{\pm}0.01$                        \\
 $q$                                      &    $4.61{\pm}0.07$              &   $9.02{\pm}0.10$    &   $8.72{\pm}0.09$    &  $4.52{\pm}0.14$     &    $6.5{\pm}0.6$        &   $4.8{\pm}1.1$                          \\
 $R_{\rm in}$\,$(R_{g}={\rm GM/c^{2}})$   &    $1.36{\pm}0.07$              &   $1.244{\pm}0.009$  &   $1.314{\pm}0.008$  &  $2.09{\pm}0.06$     &    $2.45{\pm}0.08$      &   $2.30{\pm}0.20$                          \\
 $R_{\rm out}$\,$(R_{g})$                 &    $400$                        &   $400$              &   $400$              &  $400$               &    $400$                &   $400$                                    \\
 $i$\,$(^{\circ})$                        &    $44.0{\pm}0.7$               &   $46.2{\pm}0.4$     &   $40.31{\pm}0.19$   &  $50{\pm}10$         &    $55.7{\pm}0.5$       &   $59.5{\pm}0.5$                   \\
 Fe/solar                                 &    $7.1{\pm}0.3$                &   $7.7{\pm}0.5$      &   $9.6{\pm}0.4$      &  $4.98{\pm}0.12$     &    $4.06{\pm}0.12$      &   $4.5{\pm}0.8$                   \\
 ${\xi}$\,$(=4{\pi}F_{\rm irr}/n)$        &    $2\,155{\pm}12$              &   $2\,531{\pm}13$    &   $1\,789{\pm}14$    &  $1\,673{\pm}19$     &    $49.4{\pm}0.8$       &   $29{\pm}6$                 \\
 $F_{\rm powerlaw}/F_{\rm total}$ \footnote{Fluxes calculated in the 0.05--100\,keV energy band.}   &    $0.32$                       &   --  \footnote{For NGC~1313 X--2 and M~81 X--6 the best fits were obtained considering fully {\tt reflection dominated} models, i.e. $F_{\rm powerlaw}/F_{\rm total}=0$.}               &   --                 &  $0.77$              &    $0.75$               &   $0.92$                                \\
 ${\chi}^{2}/{\nu}$                       &    $975/919$                    &   $991/952$          &   $956/840$          &  $1\,292/1\,275$     &    $736/577$            &   $945/779$                              \\
 $L_{\rm (0.05-100)\,keV}$\,$({\rm erg\,s^{-1}})$    &     $1.4{\times}10^{40}$  &  $2.1{\times}10^{40}$ &  $1.9{\times}10^{40}$ & $3.6{\times}10^{40}$ &    $6.7{\times}10^{40}$ &  $1.4{\times}10^{41}$                  \\
\hline
 $a$\,($=cJ/GM^{2}$)                      &    $0.960-0.998$                  &   $0.970-0.998$      &   $0.993-0.998$      &   $0.95{\pm}0.17$    &    $0.70_{-0.05}^{+0.12}$  \footnote{{\tt kerrconv} gave a worse fit (755/578) that {\tt kdblur} for this source.}   &    $0.47{\pm}0.12$ \footnote{{\tt kerrconv} provides a fit of the same quality (945/779) as {\tt kdblur} (see text). }       \\
\hline
\end{tabular}
\end{minipage}
\end{table*}

\section{Discussion}   \label{discuss}

Remarkably, a relativistically-blurred, reflection-dominated model
gives a very good fit to the sample of ULX with best quality (100\,ks
of time exposure) {\it XMM-Newton} data.  Very skewed iron line
profiles have been found, implying that the emission region is very
close to the black-hole and this fact allowed us to determine their
spin.  We find that all the ULX in our sample are close to maximally
rotating, with the exception of NGC~5408 X--1 and HOLM~II X--1, for
which the steep powerlaw did not allow us to properly determine
parameters from the Fe line. We have found that, the sources showing a
{\it reflection dominated} spectrum (NGC~1313 X--1, NGC~1313 X--2 and
M~81 X--6) have maximally spinning black-holes.  This fact can be
understood if light bending is an important effect for these
sources. For these sources, the high-energy source (i.e. the jet or
corona) is very close to the black hole and the observed direct
continuum (i.e. powerlaw) is very low. This would correspond to {\tt
  Regime I} of \citet{miniutti04}, corresponding to a low height of
the primary source, where strong light suffered by the primary
radiation dramatically reduces the observed powerlaw emission
component at infinity. HOLM~IX X--1 shows a {\it powerlaw dominated} 
spectrum with a slightly narrower Fe line profile. This would 
correspond to {\tt Regime II} of \citet{miniutti04}, when the height of the
primary source is larger, and the observed direct continuum
(i.e. powerlaw) stronger. The line profile is narrower than in regime
I because the emissivity profile of the disk is flatter, as result
of a nearly isotropic illumination. For NGC~5408 X--1 and HOLM~II X--1,
we find a clear reflection component but the steepness of the
spectra means that the Fe K features are less well-measured.

The spectral solution we have found here for ULX suggests a new
explanation for accretion onto spinning black holes. Our model assumes
that the dominant source of radiation is a power-law continuum
produced a few gravitational radii above a rapidly spinning black
hole.  Little thermal radiation is produced by the disc and is
undetected by the current data. We assume that the power for this
source is extracted magnetically from the disc and transferred to the
emission region by magnetic fields. Some of the power may even be
extracted magnetically from the spin of the black hole \citep{blandford77}.
Since radiation is only produced above the disc,
radiation pressure need not oppose accretion. Indeed it will help
squash the disc and maintain the high surface density required for our
relatively low ionization parameters and thus observable reflection.
(Note that strong light bending occurs in the region being considered
which is very close to a black hole, meaning that there is beaming but
in the {\emph opposite} sense to that normally envisaged.)

The relevant radiation for computing the physical Eddington limit in
this situation is the thermal disc radiation, not the power-law
continuum and reflection associated with it. Since we detect no such
thermal radiation, then the situation may be sub-Eddington, even for
stellar mass black holes. Whether this solution can work depends on
the extent to which magnetic energy extraction can be clean, in the
sense of not requiring considerable thermal energy release. We note
that the accretion flow is super-Eddington in the conventional
interpretation in which the total energy release is considered
(especially since some radiation falls straight into the black hole).

If the above solution is appropriate for these objects then they can
either be stellar mass black holes with masses of ${\approx}10\Msun$ or
IMBH of $100$s$\Msun$. Their spin is then likely to be natal. If they 
form through stellar collapse and the progenitor star was highly 
spinning, then it would be likely that the black hole would remain 
highly spinning. Many studies have argued that a binary companion 
can spin up a massive star but the magnitude
of the spin up is still a matter of debate (e.g.,
\citealt{paczynski98,fryer99,brown00}). Accretion would presumably
come from a binary companion. An interesting alternative is that they
accrete from a debris disc produced by fallback from the (failed)
supernova or collapsar \citep{li03}. This could account for the high
metallicity we infer (note that hydrogen-poor accretion would also
slightly raise the Eddington limit).

In summary, some ULX may consist of fast spinning black holes
accreting rapidly from a metal-rich disc. Accretion power
passes magnetically to the primary emission region which lies just
outside the accretion flow. Strong reflection occurs, creating a soft
excess and very broad iron-K features in the 0.3-10 keV X-ray
spectrum. Sensitive hard X-ray spectra in the 10--50~keV band may test
this scenario.

\section*{Acknowledgments}
This work is based on observations made with  XMM-Newton, an
ESA science mission with instruments and contributions directly
funded by ESA member states and the USA (NASA). We thank 
the anonymous referee for helpful comments.

\end{document}